%% file: main.tex
\documentclass[conference]{IEEEtran}
\IEEEoverridecommandlockouts

\usepackage{cite}
\usepackage[numbers]{natbib}
\usepackage{amsmath,amssymb,amsfonts}
\usepackage{algorithm}
\usepackage{algpseudocode}
\usepackage{graphicx}
\usepackage{textcomp}
\usepackage{float}
\usepackage{xcolor}
\usepackage{array}
\usepackage{adjustbox}
\usepackage{tabularx}
\usepackage[caption=false,font=footnotesize]{subfig}
\usepackage[font=small,labelfont=bf,labelsep=colon]{caption}

\usepackage{tikz}
\usepackage{pgfplots}
\pgfplotsset{compat=1.18}
\usepgfplotslibrary{groupplots,fillbetween}
\usetikzlibrary{positioning,calc,fit}

\usepackage[hidelinks]{hyperref}

\newcolumntype{L}[1]{>{\raggedright\arraybackslash}p{#1}}
\newcolumntype{R}[1]{>{\raggedleft\arraybackslash}p{#1}}

\makeatletter
\newenvironment{breakablealgorithm}
  { % begin
   \begin{center}
     \refstepcounter{algorithm}
     \hrule height.8pt depth0pt \kern2pt
     \renewcommand{\caption}[2][\relax]{%
       {\raggedright\textbf{Algorithm~\thealgorithm} ##2\par}%
       \ifx\relax##1\relax 
         \addcontentsline{loa}{algorithm}{\protect\numberline{\thealgorithm}##2}%
       \else
         \addcontentsline{loa}{algorithm}{\protect\numberline{\thealgorithm}##1}%
       \fi
       \kern2pt\hrule\kern2pt
     }
  }
  { % end
     \kern2pt\hrule\relax
   \end{center}
  }
\makeatother

\begin{document}

\title{RIS-Assisted D-MIMO for Energy-Efficient 6G Indoor Networks\\
\thanks{This work was supported by the European Commission through the Horizon Europe/JU SNS project Hexa-X-II (Grant Agreement no. 101095759), and the Advanced Digitalization program at the WiTECH Centre DisCouRSe financed by VINNOVA, Chalmers, Ericsson, Qamcom, RISE, SAAB and Volvo Cars.}
\vspace{-3mm}}

\author{
\IEEEauthorblockN{
Akshay Vayal Parambath$^{1}$,
Jose Flordelis$^{2}$,
Venkatesh Tentu$^{1}$,
Charitha Madapatha$^{1}$,\\
Fredrik Rusek$^{2}$,
Erik Bengtsson$^{2}$,
Tommy Svensson$^{1}$
}
\IEEEauthorblockA{
$^{1}$Department of Electrical Engineering, Chalmers University of Technology, Gothenburg, Sweden\\
$^{2}$Sony Europe, Lund, Sweden
}\vspace{-9mm}
}
\maketitle

\begin{abstract}
We propose an alternating optimization framework for maximizing energy efficiency (EE) in reconfigurable intelligent surface (RIS) assisted distributed MIMO (D-MIMO) systems under both coherent and non-coherent reception modes. The framework jointly optimizes access point (AP) power allocation and RIS phase configurations to improve EE under per-AP power and signal-to-interference-plus-noise ratio (SINR) constraints. Using majorization-minimization for power allocation together with per-element RIS adaptation, the framework achieves tractable optimization of this non-convex problem. Simulation results for indoor deployments with realistic power-consumption models show that the proposed scheme outperforms equal-power and random-scatterer baselines, with clear EE gains. We evaluate the performance of both reception modes and quantify the impact of RIS phase-shift optimization, RIS controller architectures (centralized vs. per-RIS control), and RIS size, providing design insights for practical RIS-assisted D-MIMO deployments in future 6G networks.
\end{abstract}

\vspace{-2mm}
\section{Introduction} 

\IEEEPARstart{A}{pproaching} the sixth generation (6G) of wireless communications, the applications and use cases are rapidly expanding, leading to a sharp rise in the number of connected devices. Although macro base-station deployments act as the backbone of wireless networks, they often struggle to meet user requirements and fairness as the network grows. To balance these demands, more sophisticated architectures are needed to complement or extend existing infrastructure while ensuring efficient and sustainable operation \cite{interdonato2019ubiquitous}. 

Distributed MIMO (D-MIMO), also known as cell-free massive MIMO (CF-mMIMO), has emerged as a promising paradigm, characterized by its adaptability and user-centric (UC) serving capability \cite{bjornson2020scalable}. In such networks, distributed access points (APs) are dynamically clustered based on user positions, or channel conditions, leveraging either estimated or measured channel state information (CSI). These UC clusters enable flexible AP coordination, improving key performance indicators such as throughput, coverage, and fairness. Despite these advantages, D-MIMO systems still suffer from severe coverage degradation, especially in high-blockage indoor environments \cite{di2020smart}. Increasing AP density could mitigate this, but it is neither cost-efficient nor energy-efficient.

Even with distributed multi-antenna APs, indoor D-MIMO remains constrained when direct AP-UE links are blocked or subject to severe pathloss, as beamforming cannot bypass obstacles \cite{di2020smart}. Reconfigurable intelligent surfaces (RISs) address this by enabling controllable reflections that form alternative paths without deploying additional APs \cite{huang2019reconfigurable}. Unlike network-controlled repeaters that rely on power-hungry amplify-and-forward relaying, RISs are nearly passive arrays of programmable reflecting elements. By intelligently configuring their phase shifts, they align reflected signals with AP-UE links, thereby enhancing signal strength, suppressing interference, and improving link reliability at minimal power cost. With minimal hardware complexity and ease of deployment, RISs complement D-MIMO and serve as an energy-efficient alternative to AP densification in indoor environments.

An additional challenge lies in the non-coherent (NC) reception inherent in D-MIMO systems, where signals from multiple APs may arrive phase-misaligned, disrupting coherent (C) combining and necessitating refined models that jointly optimize RIS configuration and power allocation under both C and NC reception. Here, coherent operation assumes that the serving APs coordinate their transmissions in both time and phase, enabling coherent signal combining at the UE. 

Several approaches have attempted to address NC reception in related architectures. For instance, mixed C and NC transmission strategies have been proposed to mitigate NC reception in CF-mMIMO-OFDM systems \cite{li2024asynchronous}. In addition, rate-splitting methods \cite{zheng2023asynchronous}, have been explored to improve SE in NC conditions. On the EE side, joint AP activation and power allocation for energy-efficient CF-mMIMO networks has been studied in \cite{van2020joint}. These works highlight the challenges of NC reception and the importance of SE/EE optimization. However, most existing works either focus on NC reception in CF-mMIMO without RIS \cite{li2024asynchronous}, \cite{zheng2023asynchronous} or consider EE optimization without jointly accounting for RIS-assisted C and NC reception \cite{van2020joint}. Our previous work \cite{parambath2024integrating} demonstrated the potential of RIS-assisted D-MIMO for improving indoor coverage and UC clustering. In that study, we showed that dynamic AP clustering combined with RIS assistance can significantly enhance signal-to-interference-plus-noise ratio (SINR), EE, and coverage probability. We further revealed that RISs enable smaller AP clusters to achieve performance comparable to larger clusters, highlighting RIS as a scalable and energy-efficient alternative to AP densification. However, that study did not explicitly address NC reception nor formulated a joint EE optimization framework.

Motivated by these gaps, this paper develops a joint alternating optimization (AO) of RIS phase-shifts and transmit powers to maximize global EE in RIS-assisted D-MIMO systems. Unlike prior works, the proposed framework evaluates the system under both C and NC reception while optimizing transmit power and RIS phase shifts for EE. A detailed power-consumption model is incorporated, including RIS controller and hardware power consumption, and the resulting EE serves as an upper-bound benchmark for practical implementations. Specifically, we incorporate 3GPP Indoor Hotspot (InH) channel models \cite{ETSI2020}. To address the resulting non-convexity, the proposed AO framework combines majorization-minimization (MM)-based power allocation with RIS phase-shift adaptation. Through simulations, we benchmark the proposed approach against equal-power and random-scatterer RIS baselines, providing insights into the roles of RIS optimization, controller architectures, and RIS size under both reception modes. Overall, the findings show that RIS-assisted D-MIMO effectively balances SE and EE, offering valuable guidance for future 6G wireless networks.
\vspace{-2.5mm}
\section{System Model}

\subsection{Network Topology and Associations}
\vspace{-5mm}
\begin{figure}[H]
    \centering
    \includegraphics[width=0.95\linewidth,height=0.42\linewidth]{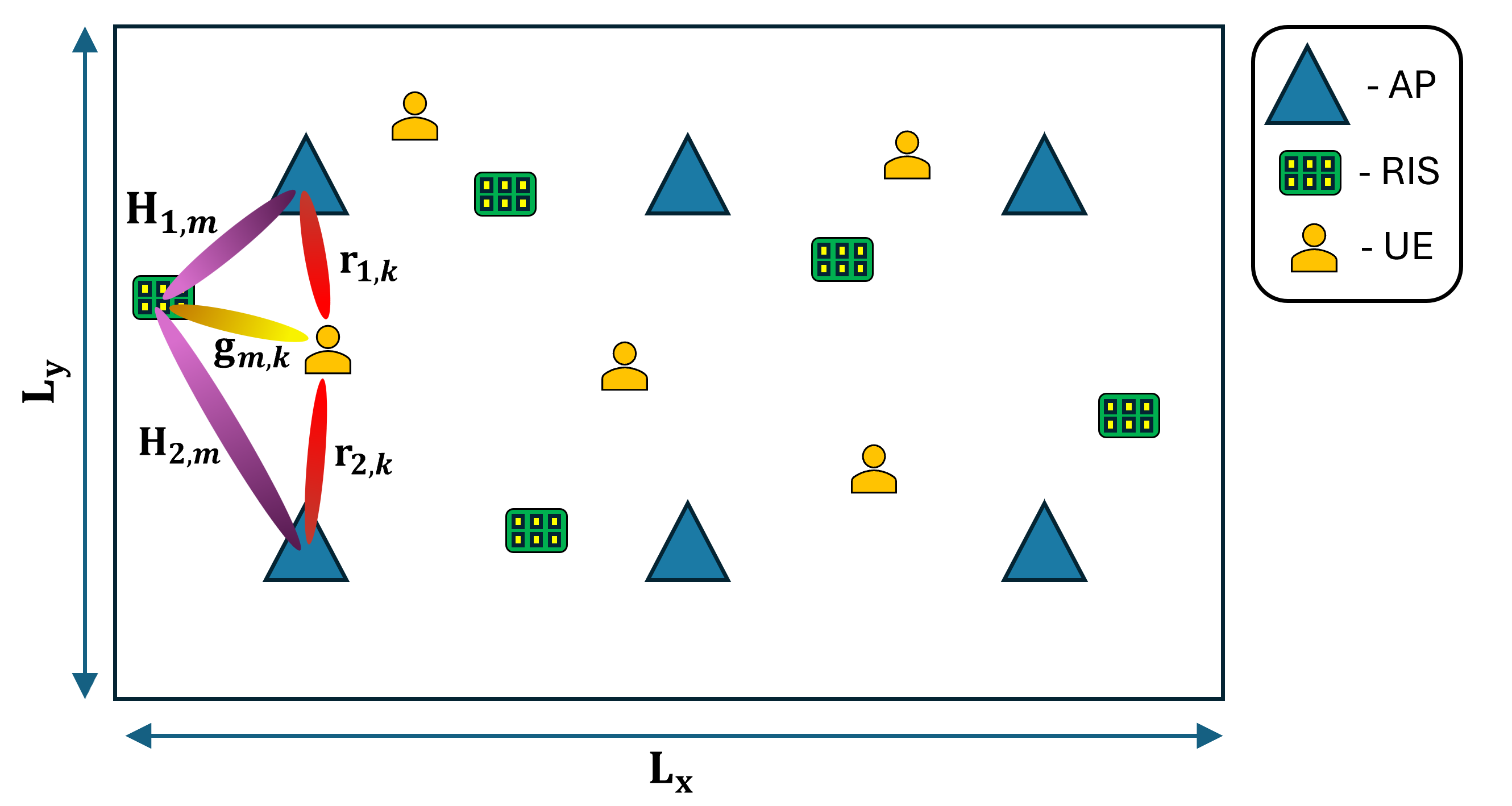}
    \vspace{-1mm}
	\caption{Illustration of an RIS-assisted D-MIMO downlink system.}
    \label{fig:System mdel}
\end{figure}
\vspace{-4.5mm}
We consider a narrowband RIS-aided D-MIMO downlink with $L$ distributed APs, $K$ single-antenna user equipments (UEs), and $M$ RISs as illustrated in Figure~\ref{fig:System mdel}. The APs are geographically distributed and connected via fronthaul links to a central processing unit (CPU) for UC coordination, joint precoder design, and RIS control. The deployment follows the 3GPP InH model \cite{ETSI2020}, with APs uniformly placed on a grid with spacing $D$ and offset $D/2$ from the walls. The grid size $(L_{\mathrm{x}}, L_{\mathrm{y}})$ and spacing $D$ are scenario dependent. The system operates at a carrier frequency $f_c$ and bandwidth $B$. The UEs are distributed following a finite homogeneous Poisson point process (FHPPP). Only a fraction of the $K$ UEs are active per scheduling slot, reflecting realistic indoor traffic loads (e.g., factory environments). Since AP locations are fixed, RIS panels are randomly deployed across the coverage area to support the spatially varying active UEs and provide additional reflected paths in blocked regions. Each AP can serve up to $N_{\text{slot}}$ UEs per downlink time slot. 

Each active UE $k \in \mathcal{K}_{\text{act}}$ is jointly served by its two strongest APs, forming the cluster $\mathcal{S}_k=\{n_1(k),n_2(k)\}\subseteq\mathcal{L}=\{1,2,\dots,L\}$. An AP cluster size of two is used to keep the system model tractable while focusing on power allocation and RIS phase optimization. In addition, each UE is assisted by one RIS selected from $\mathcal{M}=\{1,2,\dots,M\}$. The AP-UE clustering \cite{bjornson2020scalable,ye2016user}, and UE-RIS association are based on large-scale fading and geometry, so that each UE selects the RIS offering the strongest cascaded AP-RIS-UE channel. The RIS size $N_{\mathrm{RIS}}$ determines the passive beamforming gain; under ideal alignment, the reflected-signal power scales with $N_{\mathrm{RIS}}^2$, thereby improving both SE and EE.
\vspace{-1.5mm}
\subsection{Channel Model}
\vspace{-1mm}
We model each AP-UE, and AP-RIS-UE channel according to the 3GPP InH model~\cite[Table~7.4.1-1, 7.4.2-1]{ETSI2020},~\cite{wang2018applicability} as a Rician fading channel, which accounts for line-of-sight (LoS) probability, pathloss ($\mathrm{PL}$), and shadowing in indoor scenarios. For a generic MIMO channel between a transmitter $t$ with $N_t$ antennas and a receiver $r$ with $N_r$ antennas:
\vspace{-1mm}
\begin{equation}
\mathbf{H}_{t,r} = \sqrt{\beta_{t,r}}\left(
    \sqrt{\tfrac{\kappa_{t,r}}{\kappa_{t,r}+1}}\, \mathbf{a}_{r}(\theta_{r})
    \mathbf{a}_{t}^{H}(\theta_{t})
    + \sqrt{\tfrac{1}{\kappa_{t,r}+1}}\, \mathbf{W_{t,r}}
\right),
\label{eq:rician-model}
\end{equation}
\vspace{-4mm}

\noindent
where $\kappa_{t,r}$ is the Rician factor \cite[Table~7.5-6]{ETSI2020}, $\beta_{t,r} = G_t G_{r} \, 10^{-\mathrm{PL}(d_{3\mathrm{D}},f_c)/10}$ is the large-scale fading (including antenna gains $G_t,G_r$),  
$\mathbf{a}_{r}(\theta_r)\in\mathbb{C}^{N_r\times 1}$ and $\mathbf{a}_{t}(\theta_t)\in\mathbb{C}^{N_t\times 1}$ are the uniform linear array (ULA) steering vectors at AoA $\theta_r$ and AoD $\theta_t$, respectively, and $\mathbf{W_{t,r}}\sim\mathcal{CN}(\mathbf{0},\mathbf{I}_{N_r}\otimes \mathbf{I}_{N_t})$ is the Rayleigh component. For a ULA with $N$ elements and half-wavelength spacing, the steering vector is given by 
$\mathbf{a}(\theta) = \frac{1}{\sqrt{N}} [1,\, e^{j\pi \sin\theta},\, \ldots,\, e^{j\pi (N-1)\sin\theta}]^{T}$, where $\theta \in \{ \theta_t, \theta_r \}$.

We denote the AP-RIS channel by $\mathbf{H}_{n,m}\in\mathbb{C}^{N_{\mathrm{RIS}}\times N_{\mathrm{AP}}}$, the RIS-UE channel by $\mathbf{g}_{m,k}\in\mathbb{C}^{N_{\mathrm{RIS}}\times 1}$, and the direct AP-UE channel by $\mathbf{r}_{n,k}\in\mathbb{C}^{1\times N_{\mathrm{AP}}}$, which will be used in the subsequent signal model formulation. We assume CSI is available at the CPU for the considered indoor scenario with static users, while CSI acquisition is not addressed here. Each RIS $m$ applies a diagonal unit-modulus phase-shift matrix $\mathbf{\Phi}_m = \mathrm{diag}\!\left(e^{j\theta_{m,1}}, \ldots, e^{j\theta_{m,N_{\mathrm{RIS}}}}\right)$, where $\theta_{m,i}\!\in\![0,2\pi)$.

\subsubsection{Coherent Case}
In this case, all distributed APs are assumed to share a common phase reference, thereby achieving network-wide synchronization. This allows their transmitted signals to be combined constructively at the UE, provided that CSI is also well aligned, i.e., the relative phase offsets $\Delta_{n,k}$ between AP $n$ and UE $k$ are effectively zero. The propagation channel between AP $n$ and UE $k$ is
\vspace{-1.5mm}
\begin{equation}
\mathbf{h}_{n,k} \triangleq \mathbf{g}_{m_k,k}^H \boldsymbol{\Phi}_{m_k} \mathbf{H}_{n,m_k} + \mathbf{r}_{n,k},
\label{eq:coh-h}
\end{equation}
\vspace{-5mm}

\noindent
where $\mathbf{r}_{n,k}$ is the direct AP-UE link and the cascaded AP-RIS-UE contributions via the serving RIS $m_k$ are phase-aligned by appropriate choice of $\boldsymbol{\Phi}_{m_k}$.  
This model represents the best-case system performance, where RIS fully exploit their passive beamforming gain and signals from multiple APs combine coherently at the UE, yielding maximum array gain. While requiring significant synchronization overhead, this coherent case serves as an upper bound against which more practical schemes can be benchmarked.
\subsubsection{Non-Coherent Case}
In distributed deployments, however, accurate phase synchronization across transmitting APs is rarely achievable. Denoting the relative phase offset between AP $n$ and UE $k$ as $\Delta_{n,k} = \theta_n - \varphi_k$, each transceiver operates with its own oscillator, resulting in independent random phase offsets that are not known to the network \cite{li2024asynchronous}. The effective channel can be expressed as 
\vspace{-2mm}
\begin{equation}
\tilde{\mathbf{h}}_{n,k} \triangleq e^{j\Delta_{n,k}}\,\mathbf{h}_{n,k}.
\label{eq:noncoh-h}
\end{equation}
\vspace{-6.5mm}

Here, the lack of phase alignment prevents coherent combining across APs, reducing the overall array gain and leading to lower SE and EE compared to the coherent case.  
\vspace{-3mm}
\subsection{Signal Model}
\vspace{-1mm}
Let $s_k$ be the unit-power symbol for UE $k$, and $p_{n,k}\!\ge\!0$ the transmit power from AP $n$ to UE $k$. The transmit signal at AP $n$ is
\vspace{-2mm}
\begin{equation}
\mathbf{x}_n=\sum_{j=1}^{K}\sqrt{p_{n,j}}\,\mathbf{q}_{n,j}s_j,
\end{equation}
\vspace{-4mm}

\noindent
where $\mathbf{q}_{n,k}\in\mathbb{C}^{N_{\mathrm{AP}}\times 1}$ denotes the precoder at AP $n$ for UE $k$. The received signal at UE $k$ is
\vspace{-2mm}
\begin{equation}
y_k^{(\chi)} \hspace{-1mm}= \hspace{-2mm}
\sum_{n\in\mathcal{S}_k}\!\sqrt{p_{n,k}}\,\mathbf{h}_{n,k}^{(\chi)}\mathbf{q}_{n,k}\,s_k
+ \sum_{j\neq k}\sum_{n\in\mathcal{S}_j}\!\sqrt{p_{n,j}}\,\mathbf{h}_{n,k}^{(\chi)}\mathbf{q}_{n,j}\,s_j\\
+ n_k,
\end{equation}
\vspace{-5mm}

\noindent
where the first term represents the desired signal for UE $k$, the second terms accounts for multi-user interference, and $n_k\!\sim\!\mathcal{CN}(0,\sigma^2_k)$ denotes the additive Gaussian noise with thermal noise variance $\sigma^2_k$. Here, we use $\chi\!\in\!\{\mathrm{C},\mathrm{NC}\}$ to denote the reception type and write the effective channel as $\mathbf{h}_{n,k}^{(\chi)}$. The instantaneous SE of UE $k$ can be expressed as 
\vspace{-2mm}
\begin{equation}
\text{SE}_k^{(\chi)}\hspace{-1mm}=\log_2\!\left(
1+\tfrac{\big|\sum\limits_{n\in\mathcal{S}_k}\sqrt{p_{n,k}}\,\mathbf{h}_{n,k}^{(\chi)}\mathbf{q}_{n,k}\big|^2}
{\sum\limits_{j\neq k}\big|\sum\limits_{n\in\mathcal{S}_j}\sqrt{p_{n,j}}\,\mathbf{h}_{n,k}^{(\chi)}\mathbf{q}_{n,j}\big|^2+\sigma^2_k}
\right).
\label{eq:SE}
\end{equation}
\vspace{-6.7mm}
\subsection{ Power-Consumption Model }
\vspace{-1.7mm}
The total network power consumption in an RIS-assisted D-MIMO system is composed of the transmit power consumed by the power amplifiers (PAs) and several static hardware-related components.  
Let $\eta_{\mathrm{PA}}\in(0,1]$ denote the PA efficiency. The dynamic transmit power required to deliver the allocated per-UE powers $\{p_{n,k}\}$ with unit-norm precoders is
$P_{\mathrm{PA}}=\frac{1}{\eta_{\mathrm{PA}}}\sum_{n=1}^{L}\sum_{k=1}^{K}p_{n,k}$. Following \cite{bjornson2017massive}, the static power contributions comprises: (i) transceiver-chain power $P_{\mathrm{TRxC}}$, comprising one local oscillator (LO) per AP and the per-antenna RF power at APs and UEs, (ii) fixed per-AP overhead $P_{\mathrm{FIX}}$, and (iii) baseband signal-processing power $P_{\mathrm{SP}}$, scaling with $B,L,N_{\mathrm{AP}},K$ and inversely proportional to computational efficiency $\eta_{\mathrm{AP\text{-}c}}$. The RIS hardware also introduces a non-negligible cost, such that the total static power consumption is
$P_{\mathrm{static}} = P_{\mathrm{TRxC}} + P_{\mathrm{FIX}} + P_{\mathrm{SP}} + P_{\mathrm{RIS}}$ with $P_{\mathrm{RIS}} = \sum_{m=1}^{M}\!\big(N_{\mathrm{RIS}}P_{\mathrm{b}} + P_{\mathrm{RIS\text{-}ctrl}}\big)$, where $P_{\mathrm{b}}$ is the per-element biasing power \cite{tentu2022uav,huang2019reconfigurable} and $P_{\mathrm{RIS\text{-}ctrl}}$ accounts for the controller consumption \cite{wang2023static}. For the considered indoor scenario with one RIS per UE ($M=K$), this expression scales linearly with $K$, and the controller term reduces to a single contribution when centralized RIS control is adopted. The overall network power consumption \(P_{\mathrm{tot}}\) is the sum of $P_{\mathrm{PA}}$ and $P_{\mathrm{static}}$. This model highlights that although RISs introduce only lower $P_{\mathrm{b}}$, their controller architecture and the AP-side signal-processing complexity can dominate in the $P_{\mathrm{static}}$, thereby strongly influencing the achievable EE.
\vspace{-2.5mm}
\section{Optimization Framework}
\vspace{-1mm}
In this section, we first formulate the global EE maximization problem for the RIS-assisted D-MIMO system. We then introduce a compact effective channel representation to simplify the SINR structure and motivate the need for problem reformulation. This leads to an EE objective with power and SINR constraints, forming the basis for the MM-based AO algorithm developed later.
\vspace{-1mm}
\subsection{Effective Channel-Based Constants}\vspace{-1mm}
To simplify the SINR analysis in \eqref{eq:SE}, we rewrite the combined AP-RIS-UE channels into compact constants.  
This isolates desired signal and interference terms in \eqref{eq:SE}, while keeping the optimization problem tractable.  
For a given UE $k$, let $\mathcal{S}_k$ denote its serving APs and $m_k$ the associated RIS.  
The effective channel from AP $n_t(k)$ to UE $k$, $t\in\{1,2\}$, combining direct and RIS-reflected paths, is \vspace{-1mm} 
\begin{equation}
\mathbf{C}_{t,k} =
\big(\mathbf{g}_{m_k,k}^{H}\mathbf{\Phi}_{m_k}\mathbf{H}_{n_t(k),m_k}
+ \mathbf{r}_{n_t(k),k}\big)\mathbf{q}_{n_t(k),k}.
\end{equation}
\vspace{-1mm}
Based on these effective channels, the total desired-signal power for UE $k$ can be expressed as
\vspace{-1mm}
\begin{equation}
A_k(p_{n,k},\mathbf{\Phi}_{m_k}) = p_{1k}\mathbf{C}^{\mathrm{des}}_{1k} + p_{2k}\mathbf{C}^{\mathrm{des}}_{2k} 
+ \sqrt{p_{1k}p_{2k}}\,\mathbf{C}^{\mathrm{des}}_{3k},
\label{eq:Ak-compact}
\end{equation}
\vspace{-5mm}

\noindent
where $\mathbf{C}^{\mathrm{des}}_{1k}=|\mathbf{C}_{1,k}|^2$,
$\mathbf{C}^{\mathrm{des}}_{2k}=|\mathbf{C}_{2,k}|^2$,
and $\mathbf{C}^{\mathrm{des}}_{3k}=2\,\Re\{\mathbf{C}_{1,k}\mathbf{C}_{2,k}^*\}$,
corresponding to the received power from each AP and their cross-term correlation. Interference from other users is modeled similarly. Let $\mathbf{D}_{\ell j,k}$ denote the effective interfering channel from AP $n_\ell(j)$ (serving UE $j$) to UE $k$. The total interference-plus-noise power is 
\vspace{-1mm}
\begin{align}
B_k(p_{n,k},\mathbf{\Phi}_{m_k}) 
&= \!\!\sum_{j\neq k}\!\big(
p_{1j}\mathbf{C}^{\mathrm{int}}_{1j,k}
+ p_{2j}\mathbf{C}^{\mathrm{int}}_{2j,k}
+ \sqrt{p_{1j}p_{2j}}\,\mathbf{C}^{\mathrm{int}}_{3j,k}
\big)\notag \\
\vspace{-1mm}
&\quad +\, \sigma_k^2.
\label{eq:Bk-compact}
\end{align}
\vspace{-6mm}

\noindent
where $\mathbf{C}^{\mathrm{int}}_{1j,k}=|\mathbf{D}_{1j,k}|^2$,
$\mathbf{C}^{\mathrm{int}}_{2j,k}=|\mathbf{D}_{2j,k}|^2$,
and $\mathbf{C}^{\mathrm{int}}_{3j,k}=2\,\Re\{\mathbf{D}_{1j,k}\mathbf{D}_{2j,k}^*\}$.
The resulting SINR for UE $k$ is \vspace{-2mm}
\begin{equation}
    \gamma_k(p_{n,k},\mathbf{\Phi}_{m_k}) = \frac{A_k(p_{n,k},\mathbf{\Phi}_{m_k})}{B_k(p_{n,k}\mathbf{\Phi}_{m_k})}.
    \label{eq:SINR}
\end{equation}

\vspace{-4mm}
\subsection{Problem Formulation}
Using the SINR expression, the global EE maximization problem is formulated as in \cite{van2020joint}.
\vspace{-3mm}
\begin{equation}
\begin{aligned}
\max_{\{p_{n,k}\},\,\{\boldsymbol{\Phi}_{m_k}\}}
& \mathrm{EE}(p_{n,k},\boldsymbol{\Phi}_{m_k})
= \tfrac{B \sum\limits_{k=1}^K \log_2\!\big(1+\gamma_k(p_{n,k},\boldsymbol{\Phi}_{m_k})\big)}{P_{\mathrm{tot}}} \\[-5pt]
\text{s.t.}\quad
& p_{n,k}\ge 0,~~ \sum_{k=1}^K p_{n,k}\le P_n^{\max},~~\forall n,k,\\
& \gamma_k(p_{n,k},\boldsymbol{\Phi}_{m_k}) \ge \gamma_k^{\min},~~ \forall k.
\end{aligned}
\end{equation}
\vspace{-5mm}

\noindent
The first constraint limits each AP’s total transmit power to $P_n^{\max}$, ensuring per-AP power feasibility, while the second enforces $\gamma_k^{\min}$ for each UE. This problem is challenging due to its fractional, non-convex objective and bilinear coupling between power allocation and RIS phases, making conventional convex methods unsuitable. Hence, we adopt an AO strategy, detailed in the next section. Unlike power-minimization problems, the SINR constraints here need not hold with equality since the EE objective inherently balances rate and power.
\vspace{-2.2mm}
\subsection{Joint Alternating Optimization Algorithm}
We tackle the EE maximization via an AO framework that alternately optimizes (i) power allocation using MM for fixed RIS phases and (ii) RIS phase shifts for fixed powers. Iterative refinement of both variables enables efficient handling of the problem’s non-convexity. The compact SINR terms $A_k(p)$ and $B_k(p)$ from \eqref{eq:Ak-compact}-\eqref{eq:Bk-compact} are employed throughout.

\subsubsection{MM-Based Power Allocation (fixed $\mathbf{\Phi}_{m_k}$)}
Given fixed RIS phases, transmit powers are optimized via MM by first convexifying the global EE objective fraction, and then applying a second MM step to approximate the SE term. At iteration $t$, auxiliary variables $y_k^{(t)}$ are introduced to locally approximate the SINR in~\eqref{eq:SINR}:
\vspace{-1.5mm}
\begin{equation}
y_k^{(t)} \triangleq \sqrt{\tfrac{A_k(p_{n,k}^{(t)},\mathbf{\Phi}_{m_k})}{B_k(p_{n,k}^{(t)},\mathbf{\Phi}_{m_k})}}, 
\label{eq:aux_vars1}
\end{equation}
\vspace{-2mm}
\begin{equation}
\begin{aligned}
\text{SE}_{q_k}^{(t)}\!\big(p_{n,k}^{(t)},\boldsymbol{\Phi}_{m_k}\big)\triangleq
\log_2\!\Big(&1+2y_k^{(t)}\sqrt{A_k(p_{n,k}^{(t)},\boldsymbol{\Phi}_{m_k})}\\[-2pt]
&-(y_k^{(t)})^2 B_k(p_{n,k}^{(t)},\boldsymbol{\Phi}_{m_k})\Big),
\end{aligned}
\end{equation}
\vspace{-3.5mm}

\noindent
which locally approximates the SINR. To address the fractional EE objective, a second MM step introduces a global auxiliary ratio $\nu_{k}^{(t)}$ that linearizes the SE-power coupling per iteration, defined as 
\vspace{-3mm}
\begin{equation}
\nu^{(t)} \triangleq \tfrac{\sum_{k=1}^{K}\sqrt{B\,\text{SE}_{q_k}^{(t)}\!\big(p_{n,k}^{(t)}, \mathbf{\Phi}_{m_k}\big)}}{P_{\mathrm{tot}}^{(t)}}.
\label{eq:aux_vars2}
\end{equation}
\vspace{-5mm}

\noindent
The ratio $\nu_{k}^{(t)}$ is then updated from the current power and rate estimates to balance rate and $P_{\mathrm{tot}}^{(t)}$ in each MM iteration.

\noindent
After the MM reformulation, the objective function remains non-convex due to the bilinear dependence of $B_k(p_{n,k}^{(t)}, \mathbf{\Phi}_{m_k})$ on the transmit powers, preventing direct optimization. Thus, it is linearized around the current iterate $p_{n,k}^{(t)}$ using a first-order Taylor approximation, yielding a convex surrogate subproblem \cite{boyd2004convex}. The resulting MM surrogate maximization problem is
\vspace{-2mm}
\begin{equation}
\begin{aligned}
\max_{\,\{p_{n,k}\}} \quad 
& 2\nu_{k}^{(t)} \sum_{k=1}^{K} \sqrt{B\, \text{SE}_{q_k}^{(t)}\!\big(p_{n,k}^{(t)}, \mathbf{\Phi}_{m_k}\big)}
\hspace{-1mm}\;-\;\hspace{-1mm} \big(\nu_{k}^{(t)}\big)^{2} P_{\mathrm{tot}}(p_{n,k}^{(t)}) \\[-1pt]
\hspace{-2.5mm}\text{s.t.}\quad 
& \hspace{-2.5mm}2y_k^{(t)}\!\sqrt{A_k(p_{n,k}^{(t)},\boldsymbol{\Phi}_{m_k})}
 \hspace{-1.5mm}\;-\;\hspace{-1.5mm} (y_k^{(t)})^2 B_k(p_{n,k}^{(t)},\boldsymbol{\Phi}_{m_k})
 \ge \gamma_k^{\min},\\[1.5mm]
& \,p_{n,k}^{(t)}\ \ge 0,\quad \hspace{-1.5mm}\forall k, \quad \sum_{k \in \mathcal{K}_\ell} p_{\ell k}^{(t)} \le P_\ell^{\max},\quad \hspace{-1.5mm}\forall \ell,
\end{aligned}
\label{surrobj}
\end{equation}
\vspace{-5mm}

\noindent
where $\nu_{k}^{(t)}$ is the global auxiliary ratio. Solving this convex subproblem yields the updated powers $p_{n,k}^{(t+1)}$, ensuring non-decreasing EE across iterations.

\subsubsection{RIS Phase Optimization (fixed powers)}
For a fixed power allocation, each RIS matrix $\boldsymbol{\Phi}_{m_k}$ assisting UE~$k$ is updated to enhance the desired signal, as shown in Algorithm~\ref{alg:AO-EE}. After each update, the effective channel constants $A_k(\cdot)$ and $B_k(\cdot)$ are recomputed, ensuring consistency with the new RIS configuration. This step exploits the passive beamforming gain of RIS while remaining computationally lightweight.

\subsubsection{Outer Loop and Stopping Rule}
The MM-based power allocation and RIS updates alternate until convergence, i.e., until the global EE improvement between two consecutive outer iterations fall below a threshold $\epsilon$, or when the maximum iteration count is reached. This stopping rule ensures monotonic EE improvement and stable convergence of the AO procedure. The full procedure is outlined in Algorithm~\ref{alg:AO-EE}.

\subsubsection{Computational Complexity}
In each outer AO iteration, the algorithm alternates MM-based power allocation and RIS phase optimization. Computing the auxiliary variables $y_k^{(t)}$ and $\nu^{(t)}$ in \eqref{eq:aux_vars1}-\eqref{eq:aux_vars2} has negligible complexity, while the dominant cost arises from solving the convex surrogate problem in \eqref{surrobj}. In the considered setup, each UE is served by two APs, giving $n=2K$ real optimization variables. The surrogate problem includes $K$ SINR, $N$ per-AP power, and $2K$ non-negativity constraints, giving $m=3K+N$ constraints. Using a standard interior-point method, the worst-case per-iteration complexity is $\mathcal{O}\!\left((n+m)^{3/2}n^2\right)
= \mathcal{O}\!\left((5K+N)^{3/2}(2K)^2\right)$. For fixed powers, RIS phase optimization uses per-element updates with linear complexity $\mathcal{O}(KM)$ per outer iteration, where $M$ is the number of RIS elements. The total complexity scales linearly with the number of outer AO iterations $T_{\text{outer}}$.
\vspace{-1mm}
\begin{breakablealgorithm}
\caption{AO Framework for EE Maximization}
\label{alg:AO-EE}
\small

\textbf{Input:} Channel gains; $\{n_1(k),n_2(k)\}$; $\{\gamma_k^{\min}\}$; $\sigma^2$; $\{P_\ell^{\max}\}$; $P_{\text{static}}$; $B$; max inner/outer iters $T,T_{\text{outer}}$; $K$.   

\hspace{-17mm}\textbf{Output:} Final power allocation $p_{n,k}^{(t+1)}$ and $\boldsymbol{\Phi}_{m_k}$.
\begin{algorithmic}[1]
\State \textbf{Initialize:} Set $p_{n,k}^{(0)} = p_{n,k}^0$, $\bar{\text{EE}}^{(0)} = -\infty$
\For{$n = 1$ to $T_{\text{outer}}$} \Comment{Outer AO loop}
  \State \textbf{---MM-Based Power Optimization (fixed RIS phases)---}
  \For{$t = 0$ to $T-1$}
    \State Compute $A_k^{(t)}, B_k^{(t)}, y_k^{(t)}, \text{SEeq}_k^{(t)}$, and $\nu^{(t)}$
    \State Solve surrogate problem in Eq. \eqref{surrobj}
    \State Let $p_{n,k}^{(t+1)} \gets$ optimal solution of surrogate
    \State Update $p_{n,k}^{(t)} \gets p_{n,k}^{(t+1)}$
    \If{convergence met} \textbf{break} \EndIf
  \EndFor
  \State \textbf{---RIS Phase Optimization (fixed powers)---}
  \For{$k = 1$ to $K$}
    \State Load fixed $\{\mathbf{q}_{1,k},\mathbf{q}_{2,k}\}$ and $\{p_{1k}^{(t+1)},p_{2k}^{(t+1)}\}$
    \Repeat
      \State Compute $\mathbf{g}_k = \mathbf{g}_{1,k}^{\mathrm{H}}$, $\mathbf{h}_k = \mathbf{H}_{1,1} \mathbf{q}_{1,k} + \mathbf{H}_{2, 1} \mathbf{q}_{2,k}$
      \State Update $\boldsymbol{\Phi}_{m_k} = -\angle(\mathbf{g}_k \mathbf{h}_k)$
      \State Apply phase correction $\theta = \angle(\mathbf{r}_{1,k}\mathbf{q}_{1,k})$
      \State Update cascaded channel with $\boldsymbol{\Phi}_{m_k}$
    \Until{convergence}
    \State Save $\boldsymbol{\Phi}_{m_k}$ and Update \hspace{-1.5mm}$P_k^{\text{rx}}$ = \hspace{-1.5mm}$\sum\limits_{\ell \in \mathcal{L}_k} p_{\ell k}^{(t+1)}
\big(|\mathbf{r}_{\ell,k}\mathbf{q}_{\ell,k}|^2$ + \hspace{-1.5mm}$|\mathbf{g}_{1,k}\boldsymbol{\Phi}_{m_k}\mathbf{H}_{\ell,1}\mathbf{q}_{\ell,k}|^2 \big)$
  \State Recalculate channel gains using updated $\boldsymbol{\Phi}_{m_k}$ for UE $k$
  \EndFor

  \State Compute $\text{EE}^{(n)}$
  \If{$n>1$ and $\bar{\text{EE}}^{(n)}-\bar{\text{EE}}^{(n-1)}<\epsilon$} \textbf{break} \Comment{Converged} \EndIf

\EndFor
\end{algorithmic}
\end{breakablealgorithm}

\vspace{-2mm}
\section{Performance Evaluation}
\subsection{Simulation Setup}

We consider a $300 \times 150~\text{m}^2$ indoor area where $L$ APs are uniformly deployed following the 3GPP InH model. The system operates at $f_c$~=~4~GHz with $B$~=~20~MHz. Each AP ($N_{\mathrm{AP}}=4$) serves up to $N_{\text{slot}}$~=~4 UEs per time slot. A total of $K$~=~100 single-antenna UEs are distributed following a FHPPP, with $10\%$ active UEs per slot. Each active UE $k \in \mathcal{K}_{\text{act}}$ is associated with its two strongest APs, forming a UC cluster $\mathcal{S}_k \subseteq \mathcal{L}$. We deploy $M$~=~10 RIS panels, each with $N_{\mathrm{RIS}}$ elements, randomly placed in the area, independent of the AP grid. Each UE is assisted by one available RIS applying a diagonal unit-modulus reflection matrix, where each element-wise phase shift $\theta_{m,i} \in [0,2\pi)$. This RIS control provides an upper bound to practical codebook-based approaches.
Each AP’s transmit power is limited by $P_{\ell}^{\max}$, and $\sigma^2$ is obtained from noise power spectral density of $N_0$~=~-174~dBm/Hz over $B$. We set $\gamma_k^{\min}$~=~0 to study intrinsic SE-EE behavior without explicit quality-of-service constraints, since enforcing positive SINR targets for all users may lead to infeasibility under the considered deployment and power constraints. The proposed framework is evaluated via Monte Carlo simulations, and remaining parameters are listed in Table~\ref{tab:sim-params}.

\vspace{-3mm}
\begin{table}[H]
\centering
\caption{Simulation Parameters}
\vspace{-1.75mm}
\label{tab:sim-params}
\footnotesize
\setlength{\tabcolsep}{4pt}
\renewcommand{\arraystretch}{1.12}
\begin{adjustbox}{max width=\columnwidth}
\begin{tabular}{l c}
\hline
\textbf{Parameter} & \textbf{Value} \\ 
\hline
$L, N_{\mathrm{RIS}}, N_{\mathrm{UE}}$, $\kappa$ & 9, 256 (per RIS), 1, 5 dB \\
Antenna gains $G_{\mathrm{TX}}, G_{\mathrm{RX}}, G_{\mathrm{RIS}}$ & 5 dB, 2 dB, 4 dB \\
Shadowing $\chi_{\mathrm{LoS}}, \chi_{\mathrm{NLoS}}$ & 3 dB, 8.03 dB \\
$P_{\mathrm{FIX}}, P_{\mathrm{LO}}$ & 0.875 W, 0.1 W (per AP) \\
$P_{\mathrm{AP\text{-}ant}}, P_{\mathrm{UE\text{-}ant}}$ & 0.2 W, 0.1 W (per antenna) \\
$\eta_{\mathrm{PA}}, P_{\mathrm{b}}, P_{\mathrm{RIS\text{-}ctrl}}, \eta_{\mathrm{AP\text{-}c}}$ & 0.4, 0.997 mW, 4.8 W, 750 Gflop/s \\
\hline
\end{tabular}
\end{adjustbox}
\end{table}
\vspace{-6mm}
\subsection{Results \& Discussion}
To show the effectiveness of the proposed joint AO, we evaluate global EE and sum SE under various baselines and system settings. We study the (i) benefits of RIS integration, AP density variations, and baseline comparisons of power allocation and reception models; (ii) gain of joint transmit power and RIS phase optimization; (iii) impact of RIS controller power consumption models; and (iv) effect of RIS size.

\subsubsection{SE and EE Behavior}
\vspace{-4.5mm}
\begin{figure}[H]
  \centering
  \input{figures/tikz/RIS_effect2}
  \vspace{-6.5mm}
  \caption{Sum SE and global EE vs. $P_{\mathrm{T}}$ for a D-MIMO system, comparing benchmarks of RIS integration, and global EE optimization.}
  \label{fig:EE_SE_baseline}
\end{figure}
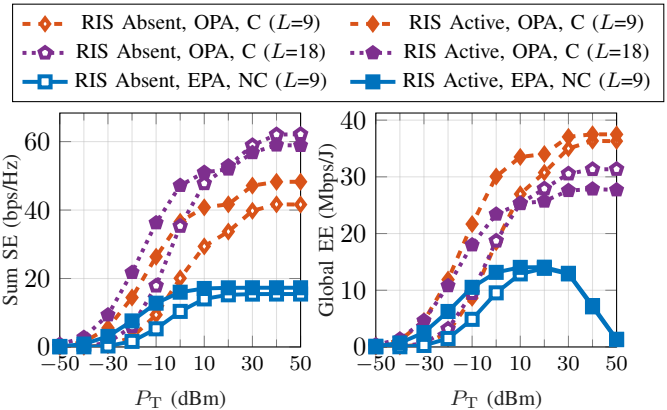
\vspace{-5.5mm}
Figure~\ref{fig:EE_SE_baseline} shows the sum SE and global EE versus $P_{\text{T}}$ \footnote{Typical indoor AP transmit powers are below $30$-$35$\,dBm. Higher transmit power values are considered here only to illustrate EE-SE trends over a wider operating range, as used in Figures~\ref{fig:EE_SE_baseline}, \ref{fig:RIS_opt}, and \ref{fig:RIS_elem}.}. First, optimized power allocation (OPA) consistently outperforms equal power allocation (EPA), particularly at medium and high $P_{\text{T}}$. The MM-based method adapts transmit power to channel conditions by favoring strong links while avoiding excess power on weaker ones once their SE contribution saturates. As a result, OPA avoids unnecessary transmit power increases after SE saturation, preserving energy and sustaining EE gains. In contrast, EPA uniformly distributes $P_{\text{T}}$ across UEs, wasting power on weaker links and causing early EE saturation or decline at high $P_{\text{T}}$. The EE gap is therefore more pronounced due to inefficient power use.

Second, RIS optimization and AP densification reveal clear performance trade-offs. When RISs are absent, both SE and EE plateau early since UEs rely solely on direct AP-UE links. Activating RISs with optimized phase-shifts (RIS-Opt) i.e., RIS Active, strengthens the effective channel, particularly at moderate $P_{\text{T}}$, where reflected paths contribute most. Increasing $L$ from $9$ to $18$ improves SE, but the added APs increase total power consumption, slightly reducing EE. At high $P_{\text{T}}$, the EE gap narrows between RIS Active and RIS Absent cases with OPA and C combining for $L = 9$, despite higher SE with RIS Active. This indicates that beyond a certain $P_{\text{T}}$, EE gains from RIS integration diminish as $P_{\mathrm{tot}}$ dominates, while higher $P_{\text{T}}$ still yield moderate SE gains from RIS and AP densification.

Third, the reception model significantly affects performance. C combining achieves the highest SE and EE by enabling coherent signal addition across APs. NC combining, while simpler to implement, yields lower performance since signals add only in power domain. Nevertheless, with RIS Active, NC still outperforms the RIS Absent case, showing that RISs remain beneficial even without perfect synchronization.

\subsubsection{Joint Transmit Power and RIS Phase-shift Optimization Significance}
\vspace{-5mm}
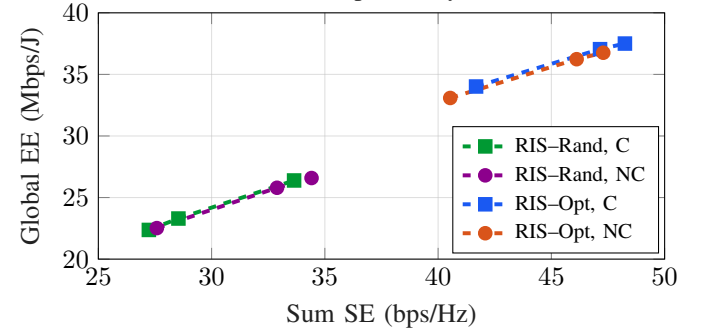
\begin{figure}[H]
    \centering
    \input{figures/tikz/RIS_opt3}
    \vspace{-6.75mm}
    \caption{Global EE vs. sum SE for a RIS-assisted D-MIMO system ($L$~=~9, $N_{\text{RIS}}$~=~256) under C and NC reception, comparing RIS-Opt and RIS-Rand with OPA across $P_{\text{T}}\!\in\![20,40]$~dBm.}
    \label{fig:RIS_opt}
\end{figure}
\vspace{-5mm}
Figure~\ref{fig:RIS_opt} shows the global EE-sum SE tradeoff for RIS-assisted D-MIMO networks with OPA at $P_{\text{T}} \in \{20,30,40\}$~dBm. The comparison with random-scatterer (RIS-Rand) baselines under both C and NC reception show that RIS-Opt provides consistent performance gains across all operating points. With RIS-Opt, reflected links constructively combine with direct AP-UE channels, enabling higher SE and translating into improved EE. In contrast, RIS-Rand results in weaker effective channels and lower EE for a given SE. Furthermore, the performance gap between C and NC reception is marginal, indicating that RIS-Opt with OPA remains effective even without coherent combining. However, in both reception models, the relative gain of RIS-Opt over RIS-Rand remains evident. These results highlight that RIS optimization is critical for unlocking the full EE potential of RIS-assisted D-MIMO systems.

\subsubsection{RIS Controller Power Consumption Models}

\begin{figure}[H]
    \centering
    \input{figures/tikz/RIS_cntrl}
    \vspace{-1.75mm}
    \caption{Global EE of a RIS-aided D-MIMO system ($L=9$, $N_{\text{RIS}}=256$) under C and NC combining for centralized and per-RIS control.}
    \label{fig:EE_ctrl}
\end{figure}
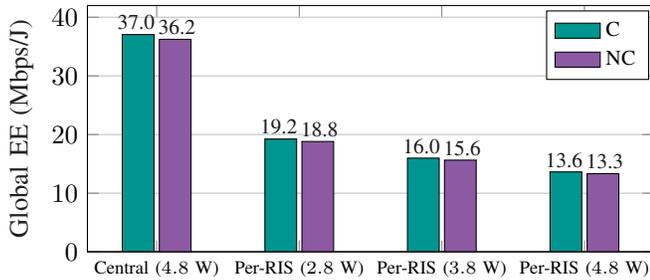

Figure~\ref{fig:EE_ctrl} evaluates the impact of RIS controller power at $P_{\mathrm{T}}$~=~30~dBm using optimized transmit power and RIS-Opt. We compare a centralized controller (shared, $P_{\mathrm{RIS\text{-}ctrl}}$~=~4.8W)  \cite{wang2023static} with per-RIS controllers having $P_{\mathrm{RIS\text{-}ctrl
}}\!\in\!\{2.8,3.8,4.8\}$~W under C and NC combining. The centralized option attains the highest EE since its overhead is effectively independent of $M$. For per-RIS control, EE decreases monotonically with $P_{\mathrm{RIS\text{-}ctrl
}}$ as the static power scales with $M$. Among per-RIS controllers, the 2.8~W case achieves the highest EE, whereas 3.8-4.8~W yield noticeably lower EE. Across architectures, the C/NC ordering is consistent, but controller power dominates EE, favoring centralized or low-power per-RIS control as $M$ increases.
\vspace{-3.8mm}
\vspace{-5mm}
\subsubsection{Effect of RIS Size}

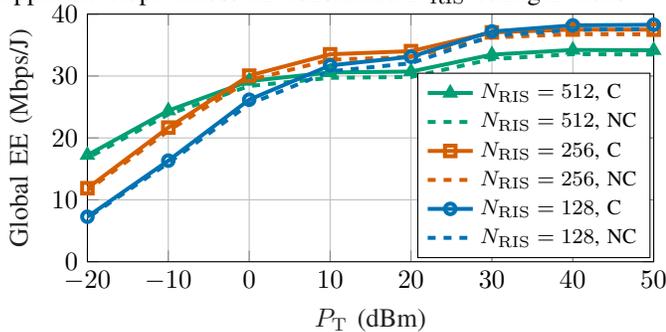
\begin{figure}[H]
    \centering
    \input{figures/tikz/RIS_element}
    \vspace{-6.75mm}
    \caption{Global EE vs. $P_{\mathrm{T}}$ for a RIS-assisted D-MIMO system with $L=9$ APs, under C and NC reception using OPA and RIS-Opt.}
    \label{fig:RIS_elem}
\end{figure}

Figure~\ref{fig:RIS_elem} illustrate that increasing $N_{\mathrm{RIS}}$ enhances EE in the low-to-moderate power range due to stronger passive beamforming, though with diminishing returns; the gain from 128 to 256 elements is evident, while the improvement from 256 to 512 is modest. As $P_{\mathrm{T}}$ increases further, the total power in the EE denominator grows faster than the achievable SE gain, leading to earlier EE saturation. As the RIS panel size increases, the higher static power outweighs the limited additional beamforming gain, resulting in only marginal benefits. At higher $P_{\mathrm{T}}$, the $N_{\mathrm{RIS}}$~=~256 case converges with the 128-element curve, while the 512-element case falls slightly below both, indicating reduced energy efficiency for excessively large panels. The C/NC ordering remains consistent, with C combining forming the upper envelope across all considered $N_{\mathrm{RIS}}$ configurations.
\vspace{-5.1mm}
\vspace{-6.8mm}
\section{Conclusion}
\vspace{-0.5mm}
This work investigated energy-efficient transmission in RIS-aided D-MIMO systems using a joint AO framework alternating MM-based power allocation and RIS phase-shift optimization. The effective-channel formulation enables efficient surrogate optimization of the non-convex EE problem. Simulation results reveal that OPA outperforms EPA, with EE peaking at moderate $P_{\mathrm{T}}$. RIS assistance improves both SE and EE, with RIS-Opt consistently outperform RIS-Rand under both combining modes, especially at moderate $P_{\mathrm{T}}$. RIS controller power is critical as centralized control achieves the highest EE, whereas per-RIS control is comparable only at low controller power; higher per-RIS budgets degrade EE as $M$ grows, emphasizing the importance of hardware-aware system design. Scaling $N_{\mathrm{RIS}}$ improves EE with diminishing returns, as most gains are captured by medium size panels, and at high $P_{\mathrm{T}}$ very large panels add only marginal EE. Finally, C combining forms the upper bound, while NC combining remains effective with RIS phase optimization and adaptive power control.

Overall, the results highlight that both signal processing algorithms and hardware architecture play pivotal roles in determining the EE of RIS-aided D-MIMO networks. Future work may extend the proposed framework to wideband systems to further assess the scalability of energy efficient D-MIMO architectures. Such an extension introduces challenges related to frequency-selective channels, RIS bandwidth limitations, and increased channel estimation and synchronization overhead, which may impact overall energy efficiency.
\vspace{-0.5mm}
%\appendices
%\section{Auxiliary Definitions for MM Power Allocation}
%\label{app:mm_aux}
%
%The auxiliary variables in the MM iterations are defined as:
%\begin{equation}
%y_k^{(t)} \triangleq \sqrt{\tfrac{A_k(p^{(t)})}{B_k(p^{(t)})}}, \quad
%\nu^{(t)} \triangleq \tfrac{\sum_{k=1}^{K}\sqrt{B\,\mathrm{SE}_{q_k}^{(t)}}}{P_{\mathrm{tot}}^{(t)}}.
%\label{eq:aux_vars}
%\end{equation}

%where $y_k^{(t)}$ approximates the SINR, and $\nu^{(t)}$ denotes the global ratio balancing rate and power consumption.

{\footnotesize
\bibliographystyle{IEEEtran}
\bibliography{IEEEabrv,references}

}

\end{document}

%% file: figures/tikz/RIS_effect2.tex
\definecolor{mycolor1}{rgb}{0.12941,0.12941,0.12941}%
\definecolor{mycolor2}{rgb}{0.00000,0.44700,0.74100}%
\definecolor{mycolor3}{rgb}{0.85000,0.32500,0.09800}%
\definecolor{mycolor4}{rgb}{0.92900,0.69400,0.12500}%
\definecolor{mycolor5}{rgb}{0.49400,0.18400,0.55600}%
\definecolor{mycolor6}{rgb}{0.46600,0.67400,0.18800}%
\pgfplotsset{compat=1.18} % or 1.17/1.16 if your TeX is older
\usepgfplotslibrary{groupplots}
\centering
\begin{tikzpicture}
\begin{groupplot}[
  group style={
    group size=2 by 1,          % 2 columns, 1 row
    horizontal sep=1cm        % gap between the two plots
  },
  width=0.54\linewidth,
  height=0.53\linewidth,
  xmin=-50, xmax=50,
  xtick={-50,-30,-10,10,30,50},  % show labels only here
  ymajorgrids,                  % (you already had this; keep it)
  yminorgrids,                  % optional
  xmajorgrids,                   % grid at the labeled ticks
  xminorgrids,                   % plus grid at minor ticks
  minor x tick num=1,            % 1 minor between majors → every 10
  scaled x ticks=false,
  xticklabel style={
  font=\footnotesize,
  /pgf/number format/.cd,
  fixed,
  precision=0
},
  grid style={opacity=0.5},
  xlabel style={font=\color{mycolor1}\footnotesize},
  ylabel style={yshift=-6pt,font=\color{mycolor1}\footnotesize},
  legend to name=commonlegend,      % collect legend entries from both plots
  legend columns=2,
  legend style={/tikz/every even column/.append style={column sep=6pt}, font=\footnotesize}
]

% ========= LEFT PANEL =========
\nextgroupplot[
  ylabel={Sum SE (bps/Hz)\footnotesize},
  xlabel={$P_{\mathrm{T}}$ \textrm{(dBm)}\footnotesize},
  ymin=0
]
% ---- paste ALL left-plot series here (unchanged) ----

\addplot [color=mycolor3, dashed, line width=1.5pt, mark size=2.2pt,
          mark=diamond*, mark options={solid, fill=white, draw=mycolor3}]
coordinates {
(-50,0.00684186822726434)
(-40,0.0621163042899094)
(-30,0.505891655555501)
(-20,2.81756531452039)
(-10,9.39104472326228)
(0,20.0682330546421)
(10,29.3870273443633)
(20,33.7116760642341)
(30,39.7769674521727)
(40,41.6756093606051)
(50,41.609868618036)
};
\addlegendentry{RIS Absent, OPA, C ($L$=9)}

\addplot [color=mycolor3, dashed, line width=1.5pt, mark size=2.2pt,
          mark=diamond*, mark options={solid, fill=mycolor3, draw=mycolor3}]
coordinates {
(-50,0.276958918515263)
(-40,1.42272407038396)
(-30,5.5468854376036)
(-20,14.4487041872901)
(-10,26.3622582487537)
(0,36.566678363596)
(10,40.7875589969684)
(20,41.6759263737538)
(30,47.1489199740668)
(40,48.2470339427037)
(50,48.2638981218727)
};
\addlegendentry{RIS Active, OPA, C ($L$=9)}

\addplot [color=mycolor5, dotted, line width=1.5pt, mark size=2.2pt,
          mark=pentagon*, mark options={solid, fill=white, draw=mycolor5}]
coordinates {
(-50,0.0256836437188093)
(-40,0.190309311865846)
(-30,1.21678054000079)
(-20,5.99559239900677)
(-10,17.8595307917026)
(0,35.304456819888)
(10,47.7657096883423)
(20,52.8237088320678)
(30,58.8921242842299)
(40,62.0525652265718)
(50,62.137514370957)
};
\addlegendentry{RIS Absent, OPA, C ($L$=18)}

\addplot [color=mycolor5, dotted, line width=1.5pt, mark size=2.1pt,
          mark=pentagon*, mark options={solid, fill=mycolor5, draw=mycolor5}]
coordinates {
(-50,0.733162792310254)
(-40,2.77540260550301)
(-30,9.34974672366009)
(-20,21.7439687420274)
(-10,36.2520051796457)
(0,47.1789366352082)
(10,50.9913563583706)
(20,51.9550349487541)
(30,56.7539742225513)
(40,58.9815696303407)
(50,58.8544791430783)
};
\addlegendentry{RIS Active, OPA, C ($L$=18)}

\addplot [color=mycolor2, line width=1.5pt, mark size=2.1pt,
          mark=square*, mark options={solid, fill=white, draw=mycolor2}]
coordinates {
(-50,0.00378034454287636)
(-40,0.0362126285890648)
(-30,0.290839102834413)
(-20,1.62731759917696)
(-10,5.30424556233811)
(0,10.3909087677425)
(10,14.0422132949816)
(20,15.2421941014106)
(30,15.4416651519404)
(40,15.4644219819688)
(50,15.4667410724458)
};
\addlegendentry{RIS Absent, EPA, NC ($L$=9)}

\addplot [color=mycolor2, line width=1.5pt, mark size=2.1pt,
          mark=square*, mark options={solid, fill=mycolor2, draw=mycolor2}]
coordinates {
(-50,0.160340805303789)
(-40,0.85403010899716)
(-30,3.08704716716324)
(-20,7.6143511854549)
(-10,12.7950170673298)
(0,16.0038975026346)
(10,17.0912950851375)
(20,17.2799952862887)
(30,17.3013093908584)
(40,17.3034725562683)
(50,17.3036892008509)
};
\addlegendentry{RIS Active, EPA, NC ($L$=9)}
% ========= RIGHT PANEL =========
\nextgroupplot[
  ylabel={Global EE (Mbps/J)\footnotesize},
  xlabel={$P_{\mathrm{T}}$ \textrm{(dBm)}\footnotesize},
  ymin=0]
% ---- paste ALL right-plot series here (unchanged) ----

\addplot [color=mycolor3, dashed, line width=1.5pt, mark size=2.1pt,
          mark=diamond*, mark options={solid, fill=white, draw=mycolor3}]
coordinates {
(-50,0.00628414991737471)
(-40,0.0570528623464948)
(-30,0.464653612364415)
(-20,2.58788820984485)
(-10,8.62546713933578)
(0,18.43103005374)
(10,26.9724597840427)
(20,30.7506279315558)
(30,35.002809031319)
(40,36.3290734591317)
(50,36.2958911371306)
};
\addlegendentry{RIS Absent, OPA, C ($L$=9)}

\addplot [color=mycolor3, dashed, line width=1.5pt, mark size=2.2pt,
          mark=diamond*, mark options={solid, fill=mycolor3, draw=mycolor3}]
coordinates {
(-50,0.227693735521376)
(-40,1.16965128871546)
(-30,4.56021050604363)
(-20,11.8785735108652)
(-10,21.6728013862245)
(0,30.0600323704145)
(10,33.5079464355461)
(20,34.016827196009)
(30,37.0485362451217)
(40,37.504135970754)
(50,37.4850897577905)
};
\addlegendentry{RIS Active, OPA, C ($L$=9)}

\addplot [color=mycolor5, dotted, line width=1.5pt, mark size=2.2pt,
          mark=pentagon*, mark options={solid, fill=white, draw=mycolor5}]
coordinates {
(-50,0.0136072284562872)
(-40,0.100826125206869)
(-30,0.644651924478249)
(-20,3.17647192153093)
(-10,9.46197759349261)
(0,18.7038084386632)
(10,25.2990828043008)
(20,27.9073622304842)
(30,30.5423821969808)
(40,31.3234457240596)
(50,31.3467211983887)
};
\addlegendentry{RIS Absent, OPA, C ($L$=18)}

\addplot [color=mycolor5, dotted, line width=1.5pt, mark size=2.1pt,
          mark=pentagon*, mark options={solid, fill=mycolor5, draw=mycolor5}]
coordinates {
(-50,0.363831557130254)
(-40,1.37729172769675)
(-30,4.63980556647136)
(-20,10.7904269318329)
(-10,17.9899786549499)
(0,23.4118167146527)
(10,25.2969285588206)
(20,25.7068939911973)
(30,27.5740362326223)
(40,27.8444097098946)
(50,27.6922731845725)
};
\addlegendentry{RIS Active, OPA, C ($L$=18)}

\addplot [color=mycolor2, line width=1.5pt, mark size=2.1pt,
          mark=square*, mark options={solid, fill=white, draw=mycolor2}]
coordinates {
(-50,0.00347218786595339)
(-40,0.0332607377477469)
(-30,0.267131181026209)
(-20,1.49466447580514)
(-10,4.87181792616206)
(0,9.54292160768443)
(10,12.884505670608)
(20,13.8594121325038)
(30,12.8799241135031)
(40,7.07071236738766)
(50,1.28447111297593)
};
\addlegendentry{RIS Absent, EPA, NC ($L$=9)}

\addplot [color=mycolor2, line width=1.5pt, mark size=2.1pt,
          mark=square*, mark options={solid, fill=mycolor2, draw=mycolor2}]
coordinates {
(-50,0.131819538815535)
(-40,0.702116057900767)
(-30,2.53792598121472)
(-20,6.25991186712738)
(-10,10.5189519757428)
(0,13.1558900330265)
(10,14.0377765436913)
(20,14.0725623693728)
(30,12.9903235054486)
(40,7.3060065883519)
(50,1.36197304982132)
};
\addlegendentry{RIS Active, EPA, NC ($L$=9)}
\end{groupplot}

% Put the shared legend centered above the two plots
\node at ($(group c1r1.north)!0.5!(group c2r1.north)+(0,8mm)$) {\ref{commonlegend}};
\end{tikzpicture}

%% file: figures/tikz/RIS_opt3.tex
\definecolor{mycolor1}{rgb}{0.10000,0.35000,0.95000}%
\definecolor{mycolor2}{rgb}{0.85000,0.33000,0.10000}%
\definecolor{mycolor3}{rgb}{0.12941,0.12941,0.12941}%
\definecolor{mycolor4}{rgb}{0.00,0.60,0.20}  % green
\definecolor{mycolor5}{rgb}{0.55,0.00,0.55}  % purple

\centering
\begin{tikzpicture}

\begin{axis}[%
width=0.85\linewidth,
height=0.37\linewidth,
scale only axis,
xmin=25,xmax=50,
xlabel style={font=\color{mycolor3}},
xlabel={Sum SE (bps/Hz)},
ymin=20,ymax=40,
ylabel style={font=\color{mycolor3}},
ylabel={Global EE (Mbps/J)},
axis background/.style={fill=white},
xmajorgrids,
ymajorgrids,
grid style={opacity=0.5},
legend style={font=\footnotesize,at={(0.625,0.54)}, anchor=north west, legend cell align=left, align=left}
]

\addplot [color=mycolor4,dashed, line width=1.5pt, mark size=2pt, mark=square*, mark options={solid, fill=white, mycolor4}]
coordinates {
(27.2304482503547,22.3708664970478)
(28.5346716818669,23.29641860011)
(33.641950226172,26.3874570937162)
};
\addlegendentry{RIS--Rand, C}

\addplot [color=mycolor5,dashed, line width=1.5pt, mark size=2pt, mark=*, mark options={solid, fill=white, mycolor5}]
coordinates {
(27.5824659538466,22.5191451014538)
(32.8885450896644,25.7945981663966)
(34.4123740509209,26.5854864323411)
};
\addlegendentry{RIS--Rand, NC}

\addplot [color=mycolor1,dashed, line width=1.5pt, mark size=2pt, mark=square*, mark options={solid, fill=white, mycolor1}]
coordinates {
(41.6759263737538,34.016827196009)
(47.1489199740668,37.0485362451217)
(48.2470339427037,37.504135970754)
};
\addlegendentry{RIS--Opt, C}

\addplot [color=mycolor2,dashed, line width=1.5pt, mark size=2pt, mark=*, mark options={solid, fill=white, mycolor2}]
coordinates {
(40.5353221332841,33.0858457797458)
(46.1158125231338,36.2351570376983)
(47.2902596860243,36.7591635127266)
};
\addlegendentry{RIS--Opt, NC}

\end{axis}

\end{tikzpicture}%

%% file: figures/tikz/RIS_cntrl.tex
\definecolor{mycolor1}{rgb}{0.00000,0.59000,0.56000}%
\definecolor{mycolor2}{rgb}{0.12941,0.12941,0.12941}%
\definecolor{mycolor3}{rgb}{0.55000,0.35000,0.64000}%
\begin{tikzpicture}

\begin{axis}[%
width=0.85\linewidth,
height=0.37\linewidth,
at={(0\linewidth,0\linewidth)},
scale only axis,
bar shift auto,
xmin=0.514285714285714,
xmax=4.48571428571429,
xtick={1,2,3,4},
xticklabels={
{Central ($4.8$ W)},
{Per-RIS ($2.8$ W)},
{Per-RIS ($3.8$ W)},
{Per-RIS ($4.8$ W)}
},
xticklabel style={align=center, font=\scriptsize,  anchor=north,rotate=0},
%enlarge x=20mm limits=0.08,
%xticklabels={{Central ($P_{\mathrm{RIS\!-\!ctrl}}=4.8\,\mathrm{W}$)},{Per-RIS ($P_{\mathrm{RIS\!-\!ctrl}}=2.8\,\mathrm{W}$)},{Per-RIS ($P_{\mathrm{RIS\!-\!ctrl}}=3.8\,\mathrm{W}$)},{Per-RIS ($P_{\mathrm{RIS\!-\!ctrl}}=4.8\,\mathrm{W}$)}},
ymin=0,
ymax=42,
ylabel style={font=\color{mycolor2}},
ylabel={Global EE (Mbps/J)},
axis background/.style={fill=white},
ymajorgrids,
legend style={font=\footnotesize,at={(0.992,0.98)},legend cell align=left, align=left}
]
\addplot[
    ybar,
    bar width=12pt,
    fill=mycolor1,
    draw=black,
    area legend
] coordinates {
(1,37.0485362451217)
(2,19.2320022922276)
(3,15.9757370815911)
(4,13.6301034083993)
};

\addplot[forget plot, color=mycolor2] coordinates {
(0.514285714285714,0)
(4.48571428571429,0)
};

\addlegendentry{C}

\addplot[
    ybar,
    bar width=12pt,
    fill=mycolor3,
    draw=black,
    area legend
] coordinates {
(1,36.2351570376983)
(2,18.8143946787947)
(3,15.6320625128862)
(4,13.3377286941402)
};

\addplot[forget plot, color=mycolor2] coordinates {
(0.514285714285714,0)
(4.48571428571429,0)
};
\addlegendentry{NC}

\node[above, align=center, inner sep=0, font=\color{white!10!black} \footnotesize]
at (axis cs:0.857,37.79) {37.0};
\node[above, align=center, inner sep=0, font=\color{white!10!black}\footnotesize]
at (axis cs:1.857,19.973) {19.2};
\node[above, align=center, inner sep=0, font=\color{white!10!black}\footnotesize]
at (axis cs:2.857,16.717) {16.0};
\node[above, align=center, inner sep=0, font=\color{white!10!black}\footnotesize]
at (axis cs:3.857,14.371) {13.6};
\node[above, align=center, inner sep=0, font=\color{white!10!black}\footnotesize]
at (axis cs:1.143,36.976) {36.2};
\node[above, align=center, inner sep=0, font=\color{white!10!black}\footnotesize]
at (axis cs:2.143,19.555) {18.8};
\node[above, align=center, inner sep=0, font=\color{white!10!black}\footnotesize]
at (axis cs:3.143,16.373) {15.6};
\node[above, align=center, inner sep=0, font=\color{white!10!black}\footnotesize]
at (axis cs:4.143,14.079) {13.3};
\end{axis}

\end{tikzpicture}%

%% file: figures/tikz/RIS_element.tex
\definecolor{mycolor1}{rgb}{0.00000,0.44706,0.69804}%
\definecolor{mycolor2}{rgb}{0.83529,0.36863,0.00000}%
\definecolor{mycolor3}{rgb}{0.00000,0.61961,0.45098}%
\definecolor{mycolor4}{rgb}{0.12941,0.12941,0.12941}%
\centering
\begin{tikzpicture}

\begin{axis}[%
width=0.85\linewidth,
height=0.372\linewidth,
at={(0\linewidth,0\linewidth)},
scale only axis,
xmin=-20,
xmax=50,
xtick={-50, -40, -30, -20, -10,   0,  10,  20,  30,  40,  50},
xlabel style={font=\color{mycolor4}},
xlabel={$P_{\mathrm{T}}$ (dBm)},
ymin=0,
ymax=40,
ylabel style={font=\color{mycolor4}},
ylabel={Global EE (Mbps/J)},
axis background/.style={fill=white},
xmajorgrids,
ymajorgrids,
legend style={font=\footnotesize,at={(0.5825,0.0145)}, anchor=south west, legend cell align=left, align=left}
]
\addplot [color=mycolor3, line width=1.5pt, mark size=2pt, mark=triangle, mark options={solid, mycolor3}]
coordinates {
(-50,1.19588565756297)
(-40,3.95747973181962)
(-30,9.4008352525349)
(-20,17.217231033145)
(-10,24.4078054789444)
(0,29.1839090667976)
(10,30.5753702808933)
(20,30.720176810616)
(30,33.451863374088)
(40,34.2111294862123)
(50,34.1520516246406)
};
\addlegendentry{$N_{\mathrm{RIS}}=512$, C}

\addplot [color=mycolor3, dashed, line width=1.5pt]
coordinates {
(-50,1.16023999477958)
(-40,3.84120084936509)
(-30,9.14069221519027)
(-20,16.7377268945635)
(-10,23.7985333455097)
(0,28.3843385882125)
(10,29.7087454417575)
(20,29.8444306214826)
(30,32.7282332112453)
(40,33.5293859221192)
(50,33.4704965930761)
};
\addlegendentry{$N_{\mathrm{RIS}}=512$, NC}

\addplot [color=mycolor2, line width=1.5pt, mark size=2pt, mark=square, mark options={solid, mycolor2}]
coordinates {
(-50,0.227693735521376)
(-40,1.16965128871546)
(-30,4.56021050604363)
(-20,11.8785735108652)
(-10,21.6728013862245)
(0,30.0600323704145)
(10,33.5079464355461)
(20,34.016827196009)
(30,37.0485362451217)
(40,37.504135970754)
(50,37.4850897577905)
};
\addlegendentry{$N_{\mathrm{RIS}}=256$, C}

\addplot [color=mycolor2, dashed, line width=1.5pt]
coordinates {
(-50,0.217184587060396)
(-40,1.12543726168552)
(-30,4.40859502898887)
(-20,11.5251092666476)
(-10,21.1512322356372)
(0,29.3118684119825)
(10,32.6265126131427)
(20,33.0858457797458)
(30,36.2351570376983)
(40,36.7591635127266)
(50,36.7305613790227)
};
\addlegendentry{$N_{\mathrm{RIS}}=256$, NC}

\addplot [color=mycolor1, line width=1.5pt, mark size=2pt, mark=o, mark options={solid, mycolor1}]
coordinates {
(-50,0.0611958639759442)
(-40,0.424185550477893)
(-30,2.05696950605648)
(-20,7.26802905741709)
(-10,16.3297621167843)
(0,26.1734032447561)
(10,31.7080361160058)
(20,33.1150342871737)
(30,37.1902466277997)
(40,38.1798108184991)
(50,38.2773239427128)
};
\addlegendentry{$N_{\mathrm{RIS}}=128$, C}

\addplot [color=mycolor1, dashed, line width=1.5pt]
coordinates {
(-50,0.0581801534702544)
(-40,0.405718186847286)
(-30,1.99325508168657)
(-20,7.07367644231751)
(-10,15.9350545855252)
(0,25.4649326356081)
(10,30.7586384449814)
(20,32.0628565044199)
(30,36.4837725429973)
(40,37.4562278498079)
(50,37.5719084113639)
};
\addlegendentry{$N_{\mathrm{RIS}}=128$, NC}
\end{axis}

\end{tikzpicture}%